# A systematic review of EEG source localization techniques and their applications on diagnosis of brain abnormalities

Shiva Asadzadeh, Tohid Yousefi Rezaii*, Soosan Beheshti, *Senior Member, IEEE,* Azra Delpak, and Saeed Meshgini

*Abstract*— In recent years, multiple noninvasive imaging modalities have been used to develop a better understanding of the human brain functionality, including positron emission tomography, single-photon emission computed tomography, and functional magnetic resonance imaging, all of which provide brain images with millimeter spatial resolutions. Despite good spatial resolution, time resolution of these methods are poor and values are about seconds. Electroencephalography (EEG) is a popular non-invasive electrophysiological technique of relatively very high time resolution which is used to measure electric potential of brain neural activity. Scalp EEG recordings can be used to perform the inverse problem in order to specify the location of the dominant sources of the brain activity. In this paper, EEG source localization research is clustered as follows: solving the inverse problem by statistical method (37.5%), diagnosis of brain abnormalities using common EEG source localization methods (18.33%), improving EEG source localization methods by non-statistical strategies (3.33%), investigating the effect of the head model on EEG source imaging results (12.5%), detection of epileptic seizures by brain activity localization based on EEG signals (20%), diagnosis and treatment of ADHD abnormalities (8.33%). Among the available methods, minimum norm solution has shown to be very promising for sources with different depths. This review investigates diseases that are diagnosed using EEG source localization techniques. In this review we provide enough evidence that the effects of psychiatric drugs on the activity of brain sources have not been enough investigated, which provides motivation for consideration in the future research using EEG source localization methods.

## Introduction

Electro- or magnetoencephalography can be used for non-invasive studies of the brain electrical activity. Scalp potential differences of the electric field driven by the neural currents are measured using EEG. The neural currents and Ohmic volume currents are driven by the electric field generate the magnetic field outside the head. MEG can measure this magnetic field [1, 2]. Only a small part of the electric field can reach the scalp sensors due to the low conductivity of the skull. The mapping from scalp sensors to brain sources is not unique. Thus, electroencephalography (EEG) that prevents cannot widely be used as an imaging modality for studying the functioning brain[3]. Unlike other brain imaging modalities such as functional magnetic resonance imaging (fMRI), positron emission tomography (PET), or functional near-infrared spectroscopy (fNIRS), the high temporal resolution of EEG allows the real-time study of brain functionality. As EEG scalp topography can be made by the arrangement of currents in the brain, solving an ill-posed inverse problem is necessary for EEG source imaging (ESI) [4]. During recent decades, brain source modeling by EEG has been an active area of research. In clinical applications, non-invasive localization of the active sources in the brain can be used to diagnose pathological, physiological, mental, and functional abnormalities related to the brain [5]. Real-time source estimates can be used to improve real-time predictions of subject's intentions compared to sensor-based predictions in applications involving brain-machine interfaces (BMI) and neurofeedback [2, 6-9]. The shape and conductivity of the skull (and scalp) strongly influence EEG signal, while this effect is less for MEG. Thus, volume conduction model and the conductivity profile of the head are necessary to estimate the sources of the measured signals.

Accurate source localization is highly dependent on the electric forward head model. The geometry and the conductivity distribution of the modeled tissue sections (scalp, skull, cerebrospinal fluid, brain grey, and white matter, etc.) are determined in the volume conduction. Magnetic resonance (MR) images of the head can provide head geometry information [10, 11]. So far, there have not been proposed any non-invasive and direct methods to measure skull and brain conductivities [12, 13]. As mentioned above, different sets of neural current sources lead to the magnetic fields and electric potentials. By considering a volume conductor model of head and appropriate source estimates (representing the activity of the neural cells), these fields can be computed. A quasistatic approximation of Maxwell's equations is used to solve the forward problem for only simple head models (typically consisting of one or more spheres) and the conductivity profile. Anisotropic, inhomogeneous and nonspherical features of real human heads affect EEG signals more than MEG. Thus, there was a need for more realistic head models. To overcome this problem, other imaging modalities, such as magnetic resonance imaging, or computerized tomography, are applied for extraction of the brain, skull, and scalp surface boundaries. This information is necessary to compute the forward problem solutions in numerical methods such as boundary element method (BEM) or finite element method (FEM) [14, 15]. In the inverse methods using the dipole source model, the sources are considered as several discrete magnetic dipoles located in certain places in a three-dimensional space within the brain. Since the electric potential at any point of the scalp can be calculated as a linear combination of the dipole amplitudes, the relationship between the potential at the scalp and the dipole amplitudes can be represented as follows [16, 17]:

$$y = Hx + e \tag{1}$$

where $x \in R^{3M}$ is related to the amplitudes of the M dipoles along the three spatial dimensions, $y \in R^N$ is the EEG data of N electrodes, the $N \times 3M$ lead field matrix H models the propagation of the electromagnetic field from the sources to the sensors[18, 19] and e is an additive white Gaussian noise.

There is no unique solution for the inverse problem. Furthermore, it needs prior knowledge about the current sources. The locations of a large number of dipoles are considered to be fixed in distributed source models (representing, and only their amplitude and orientation are calculated using MEG/EEG signals). Due to the limited number of sensors and a large number of potential source locations, the inverse problem is highly underdetermined and requires further constraints to attain a unique solution [19] (Fig.1). These assumptions might introduce further challenges such as low spatial resolution as well as errors due to localization bias in the solutions [20-22], some of which have been addressed using multiple modifications proposed in the literature.

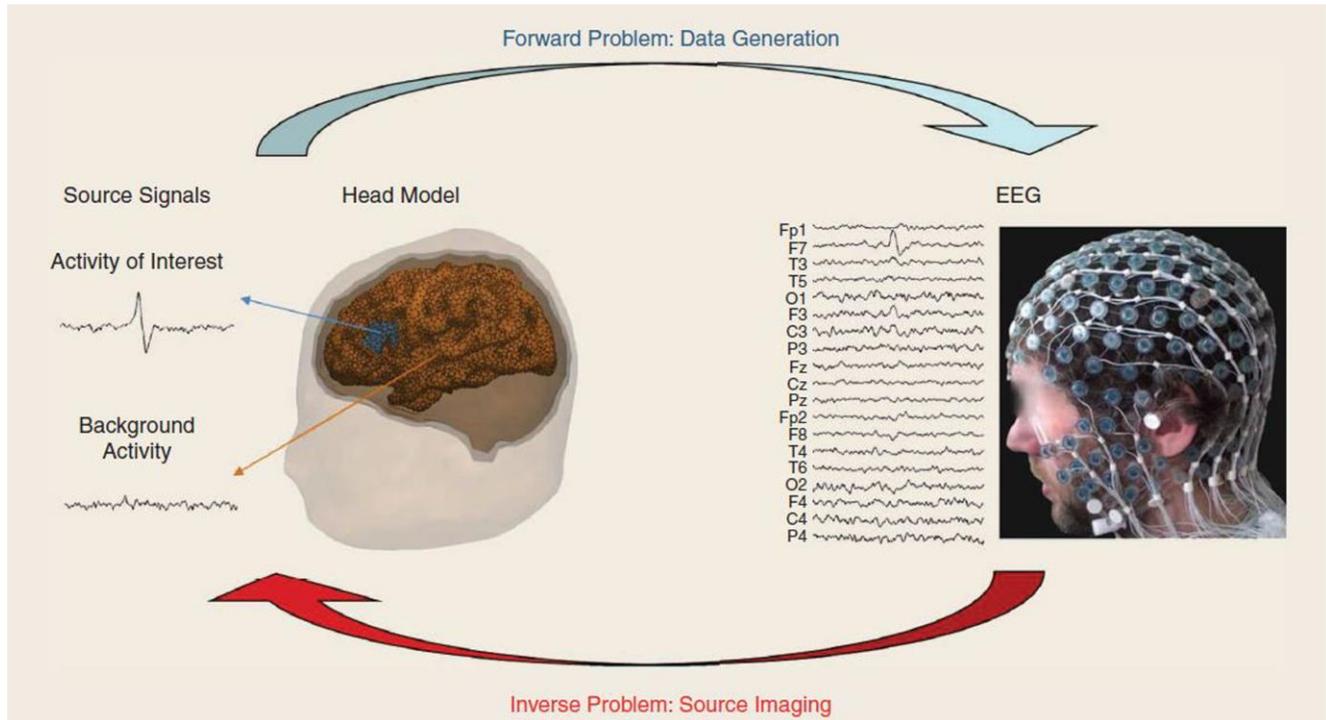

Fig.1. An illustration of the forward and inverse problems in the context of EEG[23].

During recent years, several practical techniques have been proposed to localize the sources of brain activities using EEG signals. However, there is a requirement for greater understanding of computerized EEG source localization techniques to optimize treatment and patient care in brain functional disorders. The objective of this systematic review is to identify and investigate the publications that have used source localization methods on EEG signals. It is expected that this review would aid clinical practice by informing future research and the development to diagnosis and treatment techniques in brain functional disorders.

## METHODS

In this section, research question, article selection criteria and search strategy are described.

### Research questions

The main focus of this paper is on the following research questions:
(RQ1) Which brain source localization techniques have been used to date to study brain activity sources using EEG signals?
(RQ2) Which diseases have been diagnosed and treated by the brain source localization methods so far?
(RQ3) What factors affect the accuracy of the EEG source imaging methods?
(RQ4) What are the implications for future research on brain source localization techniques in brain functional abnormalities?

### Article Selection criteria

This review paper includes studies which have focused on the following criteria: (1) presented a brain source localization method to detect brain source activities as well as related abnormalities, (2) used Electroencephalography signals, (3) demonstrated numerical and perspicuous results and (4) were written in English. Age and disease were not as limiting factor of studies.

### Search strategy

The principles in the preferred reporting items for systematic reviews and meta-analyses (PRISMA) statement is used in this review [24]. To determine the seminal works related to brain source localization techniques, a review of the literature was undertaken through a search of the following databases: PUBMED digital library, IEEE digital library, and Science Direct. Only the studies published from the year 1970 until Jan 14th, 2019 are considered. This study focuses on meeting the article selection criteria given in Section 2.2. Literature were evaluated by two independent researchers, and the agreement of both parties determined studies suitability (Fig.3).

## RESULTS

Fig.2 shows the selection process for articles included in the systematic review. A total of 180 articles were identified in the original literature search. In this paper, we are looking for researches in the field of EEG source localization. In the collected studies, 26 cases of papers were about EEG source localization using intracranial EEG and did not match the defined criteria. Also, we did not consider 34 papers that used SPECT, Computer Simulation Technology (CST), and fMRI images in their studies for this systematic review. Finally, our review consists of 120 papers.

Only twelve of the 120 studies (10%) included in this review were published by the end of 2000. Of the remaining 108 studies, 39 were published by the end of 2010 (32.5%). Finally, 69 studies (57.5%) were published between the beginning of 2011 and the end of 2018.

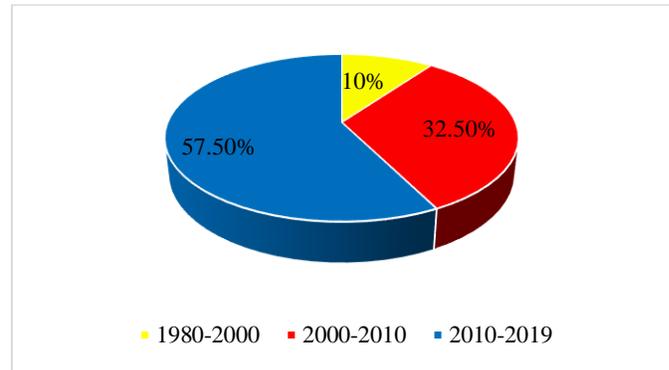

Fig. 2. Percentage of papers published from 1980 to 2000 (20-year period), 2000 to 2010 (10-year period) and 2010 to 2019 (10-year period).

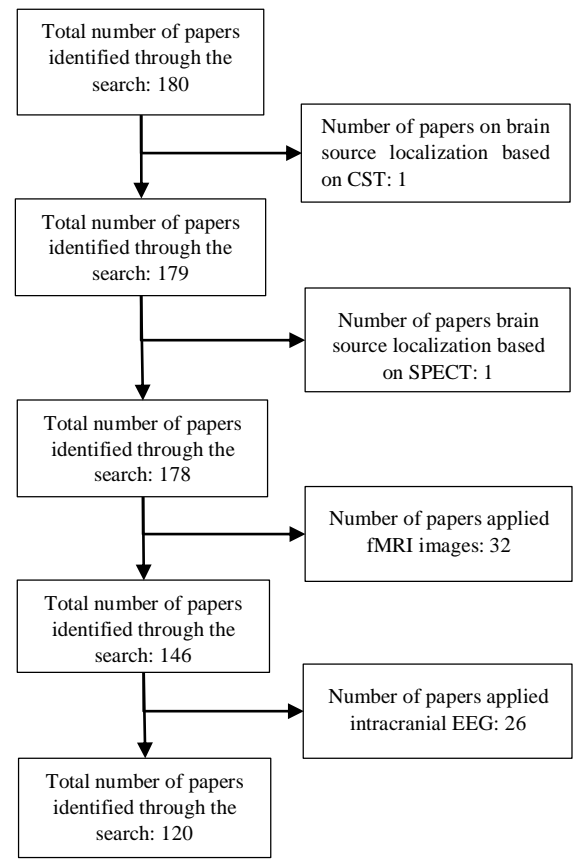

Fig.3. PRISMA diagram of the systematic review. PRISMA, Preferred Reporting Items for Systematic Reviews and Meta-Analyses[24].

On the other hand, EEG source localization studies are clustered as follows:

• Solving the inverse problem by statistical method (37.5%),

• Diagnosis of brain abnormalities using common EEG source localization methods (18.33%),

• Improving EEG source localization methods by non-statistical strategies (3.33%),

• Investigating the effect of the head model on EEG source imaging results (12.5%),

• Detection of epileptic seizures by brain activity localization based on EEG signals (20%)

• Diagnosis and treatment of ADHD abnormalities (8.33%).

In the next sections, we briefly describe the principles of these studies methods that are presented including a conceptual mind-map as depicted in Fig. 4, closing with the observed beneficial and challenging effects as a result of a fusion between the clinical and the computer science perspective.

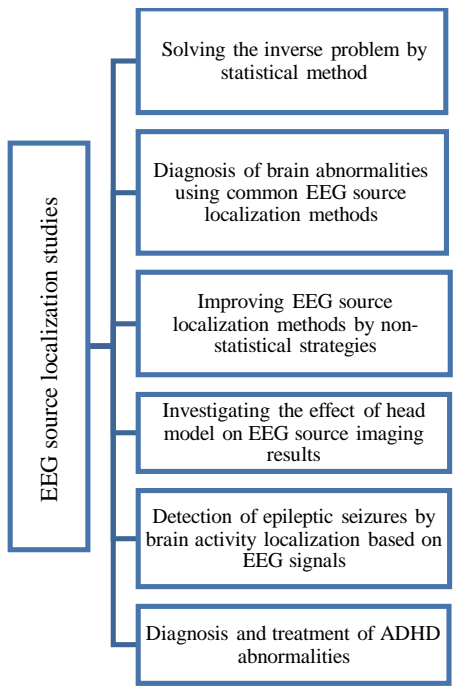

Fig. 4. Mind map of the EEG source localization studies.

*Solving the inverse problem by statistical method*

Localization of the epileptic spikes and seizures can be considered as the first reason for discussing brain sources localization. This issue was first introduced in 1982 [25]. After that, Hamalainen and Ilmoniemi presented the minimum norm method as the first mathematical method for solving the inverse problem in 1984 [26]. Although the minimum norm provides good results in terms of resolution and current estimation, it fails to address the issue of deep source localization in the outermost cortex because its solution for EEG/MEG is a harmonic function. The harmonic functions attain maximum values at the boundaries of their domain; which in this case is the outermost cortex. Furthermore, upon comparison with the newer techniques such as low-resolution electromagnetic tomography (LORETA) and WMN, the minimum norm solution has got more localization error with the disadvantage of incapability of localizing non-boundary sources [27]. Loreta is WMNE additional laplacian constraint .

Localization of the sources of EEG delta, theta, alpha, and beta frequency bands using the FFT dipole approximation was discussed in [28]. This study applies a primary brain atlas. In other words, the FFT dipole approximation produces a potential distribution for each frequency point. For each frequency point, BioLogic's DIPOLE program generates dipole sources. The potential distribution maps of the forward solutions are scaled to unity global field power [29]. The frequency of brain rhythms is very important in this method. Also, how well the computed absolute locations of the sources correspond with real brain space remains a general problem of 3-dimensional source localization procedure for momentary field distributions.

Another assumption about the brain sources is that the current density at any point of the cortex is very similar to the average current density of its neighbors. This hypothesis forms the basis of the LORETA method. LORETA provides smoother and better localization for deep sources with fewer localization errors. Disadvantages of this technique are low spatial resolution and blurred localized images of a point source with dispersion in the image[30].

It is shown that as long as the spatial patterns of the decomposition span the same signal space as the principal spatial components, the computational process of attempting to localize the sources is the same. An example is proposed using common spatial pattern decomposition and using a raw varimax rotation of a subset of the common spatial patterns. Results demonstrate that the principal component decomposition is almost useless for isolating spike and sharp wave activity in an EEG from a patient with epilepsy, while the common spatial pattern decomposition is significantly better and that the varimax rotation is better yet. That the varimax rotation is the best is shown by attempting to locate dipole sources inside the brain which account for spike and sharp wave activity on the scalp[31].

Focal underdetermined system solution (FOCUSS) is a high-resolution non-parametric technique which uses forward model that assigns a current to each element within a predetermined reconstruction region. In this algorithm, the weights are iterated at each step from the solution of the previous step. Weighted minimum norm method is applied for mathematical calculations in the recursive steps. It is reported that FOCUSS algorithm has better localization accuracy as compared with other methods and can manage non-uniquely defined localized energy sources as well as having acceptable spatial resolution. Furthermore, it is robust against non-uniquely defined localized energy sources [32].

In 1995, it was reported that neural networks are useful for solving physiological inverse problems such as EEG and MEG source localization. These methods were originally developed to localize two independent sources from EEG. Back propagation neural network (BPNN) is appropriately applied because it has the ability to install an inverse function through training using data samples [33]. However, there are very limited studies using neural networks in the field of source localization.

In 1995, recursive multiple signal classification (MUSIC) algorithm was also proposed, in which a single dipole is scanned through a grid confined to a 3D head or source volume. A signal subspace is obtained from the EEG measurements. The forward model of the dipole at each grid point is projected against this subspace. The sources are located where the projection is best onto the signal subspace. As there exist noise and error in the signal subspace and forward model, the selection of the best projection location in the practical case is an important problem in this technique. MUSIC algorithm has several limitations in terms of localizing synchronous sources. A modification of MUSIC is recursive MUSIC algorithm, which can resolve the limitations of MUSIC through the use of spatio-temporal independent topographies (IT) model [33].

A hybrid algorithm of LORETA and FOCUSS, which has the advantages of both methods, is shrinking LORETA-FOCUSS[34], although it is not validated experimentally. As a generalization of the LORETA method, sLORETA is based upon the assumption of the standardization of the current density. The spatio-temporal regularization (STR) performance of the sLORETA algorithm has been studied using a systematic comparison [35].

In this study, simple ad hoc or post hoc filtering of the data or the reconstructed current density is investigated, respectively. The results exactly show that the regularization parameters selection affects the performance of STR considerably. This assumption is considered for both the variance of the noise in the EEG measurements and the biological variance in the actual signal. eLORETA is another generalization of LORETA that focuses on reducing the localization error of deeper sources. Localization accuracy of sLORETA and eLORETA methods is better than LORETA, but their spatial resolution is not appropriate [36]. Unlike sLORETA, cortical LORETA (cLORETA) algorithm works on a surface grid [37]. Followed by sLORETA, the smallest computational complexity belongs to this algorithm [38]. Due to the low resolution of sLORETA, efforts are continued to improve this limitation. In a proposed method a post-processing algorithm called Higher Resolution sLORETA has been introduced that uses subspace based thresholding to enhance the resolution of the sLORETA source estimate. The method was compared with manual thresholding as well as automatic thresholding proposed by Otsu [39].

Brain sources are of different depths. Many methods fail to deal with deep sources. Minimum norm solutions (MNE) is one of the methods whose efficiency is proved for sources with different depths [40]. The combination of LORETA and minimum norm methods leads to hybrid weighted minimum norm method, which has a high computational complexity in time [41]. Several methods were presented to improve this constraint. Independent component analysis (ICA) was proposed for locating brain sources with acceptable accuracy. Regarding this approach, at first, principal component analysis (PCA) decomposes EEG data into signal and noise subspaces. Then, ICA is applied to the signal subspace. Due to temporally independent stationary sources, the multichannel data is separated into activation maps using ICA algorithm. It is also reported that ICA method is actually a source separation method and does not have the required performance for accurate localization [42].

A generalization of the MUSIC algorithm is spatially-extended neocortical sources MUSIC (ExSo-MUSIC). In this method, higher order statistics are applied that result in better robustness with respect to Gaussian noise distribution for unknown spatial coherence and modeling errors. Based on the computed results, the advantage of ExSo-MUSIC approach is its high performance compared to the classical MUSIC algorithms [43]. Tensor-based preprocessing can be applied to active brain source determination. In this approach, at first, space-time-frequency (STF) or space-time-wave-vector (STWV) tensor is made, and then canonical polyadic decomposition is done on its results. Compared to ExSo-MUSIC, STF-DA and STWV-DA methods have lower computational cost. This cost further increases for ExSo-MUSIC, if the number of sensors are increased. Also, increasing the number of time samples linearly raises the computational cost of ExSo-MUSIC. Therefore, it can be concluded that this algorithm is suitable for a small number of sensors and relatively large number (several thousand) of time samples [38].

A hybridization of independent component analysis (ICA) and recursively applied and projected multiple signal classification (RAP-MUSIC) were applied to dipole source estimation of epileptiform discharges. The ICA algorithm decomposes averaged EEG matrices, while RAP-MUSIC is used for source estimation. Spatial information about spikes has a high correlation with background signals. This issue leads to the low accuracy of this method [44].

In addition to ICA, source separation approaches such as blind source separation (BSS) are applied to separate various brain and extra-brain (related to artifacts) sources. These method are used as a preprocessing stage before the localization algorithms [45].

For multiple measurement vectors with constant sparsity, Novel matching pursuit (MP) based algorithms can be a handle for EEG/MEG brain source localization and estimating parameters. The advantages of such methods make it possible to reduce the residual interference inherent problems of sequential MP-based methods and or recursively applied (RAP)-MUSIC acceptably [46].

Another technique has been proposed based on the linearly constrained minimum variance and eigencanceler beamformers. A short-term estimate of the signal energy is considered as a constraint and used to select a region-of-interest (ROI). In order to map it to the brain cortex, an affine transformation is used. A complete search on the whole brain cortex leads to high computational cost. For a reduction in the computational cost, the beamforming based source localization is applied only within the ROI. Based on the results, the eigencanceler provides a more focused and less biased source [47].

The combination of sequential Monte Carlo (SMC) method and beamforming based on EEG data is used to solve the inverse problem. Estimating the coordinates of the first two non-correlated dominative brain zones is performed using the SMC method. In this approach, the beamforming method also spatially filters the EEG data [48].

In the last 10 years, Bayesian methods have become widely used to solve the inverse problem. In 2007, information about timing and spatial covariance properties of sensor data from evoked sources, interference sources and sensor noise are exploited by Bayesian method to estimate their contributions [49]. The source spatial locores and waveforms of EEG can also be calculated using multicore Berkeley packet filter (BPF).

Compared to conventional (single-core) beamforming spatial filters, the extended multicore BPF assumes clear temporal correlation among the estimated brain sources. This process is performed by suppressing activation from regions with interfering coherent sources. The hybrid multicore BPF combines the main strengths of both deterministic and Bayesian inverse problem algorithms in order to improve the localization accuracy. In this approach, prior information about approximate areas of source locations is not applied. In addition, In contrast to PF solution, the dimensionality of the problem is decreased to half using the multicore BPF [50].

Under the conditions of the range space property (RSP), the least $l_1$-norm solution is equal to at most one of the least $l_0$-norm solutions. Weighting the corresponding sensing matrix with a diagonal matrix can be used for the problem of recovering sparse signals. Thus, an $l_1$-norm minimization problem satisfying RSP can be formulated. The accuracy of this algorithm is within acceptable limits. But its runtime does not show a significant drop [51]. A $l_0 + l_1$ norm is applied by regularizing the nonzero amplitudes of the solution (by considering the solution has few non-zero elements) for source activity localization [52]. This method was proposed as an extension of Bayesian model techniques. This algorithm demonstrates better performance than the more usual $l_2$ and $l_1$ norm regularizations in terms of several evaluation criteria. In another technique, EEG sensor measurements are defined using a generative probabilistic graphical model. This model is hierarchical across spatial scales of brain regions and voxels. Then, a new Bayesian algorithm is combined with this graphical model for probabilistic inference. For sources that have a different spatial area, this algorithm provides robust reconstruction from spatially neighbor clusters of dipoles to isolated dipolar sources. The results also show that this algorithm is more robust to correlated brain activity present in real EEG data and can resolve diverse and functionally appropriate brain areas with real EEG data [53].

Currently, it is generally assumed that only a limited number of cortical regions are actually active in short periods of time. Therefore, in recent years, sparse source localization has been increasingly taken into account. The scalp measurements are summed up using few point sources, each standing for the mean activation of a close surrounding area, and providing easy interpretable visual results at destination of clinicians and neurophysiologists. Based on one assumption about main brain sources, sources of multi-channel EEG recordings may be spatially sparse, compact and smooth (SCS). In order to apply these features to the EEG inverse problem, a cortical source space covariance matrix is factorized into the multiplication of a pre-given correlation coefficient matrix using the proposed correlation-variance model; and Bayesian learning framework is employed to learn the square root of the diagonal variance matrix from the data [54]. Sparse Bayesian learning (SBL) algorithm uses an estimate of the sensor noise covariance. In this method, a good initialization of the group-sparsity profile of the sources are applied using brain atlases. Simulations show that the method is robust to the measurement noise and performs faster than other methods in the real-time. Each group of sources is considered in one region of the brain corresponding to these atlases [55].

In 2015, an original data-driven space-time-frequency dictionary was proposed, in which spatial and time-frequency sparseness are considered at the same time. Also, this technique provides smoothness in the time-frequency. Considering these hypotheses, the matching pursuit (MP) framework is used in order to choose the most appropriate atoms in this highly redundant dictionary. To reduce the computational time, the algorithm is implemented in the wavelet domain [56].

A method for the source analysis of EEG recordings is introduced as spatio-temporal unifying tomography (STOUT). Combining the sparsity constraints and an extension of the source current density into appropriate spatio-temporal basis functions remains as the foundation of this proposed method. In fact, this technique incorporates the main advantages of two available methods, namely sparse basis field expansions and time-frequency mixed-norm estimates [57]. Another method related to the sparse EEG source localization is a transfer-function-based calibration technique. This method can decrease localization error and the number of falsely recovered sources in the existence of calibration errors [58].

When covariance estimation for both source and measurement noises, linear state-space dynamics and sparsity constraints are brought together by novel computationally-efficient estimation algorithms, a very efficient localization method is created. In this method, a locally-smooth basis with sparsity performing priors is used for source covariance estimation. Furthermore, an inverse Wishart prior density is applied for EEG measurement noise covariance estimation. These model parameters are calculated by an expectation-maximization (EM) algorithm, which utilizes steady-state filtering and smoothing to accelerate computations [59].

A decomposition of the current density into a small number of spatial basis fields have been presented as another approach to solving the electro/magnetoencephalographic (EEG/MEG) inverse problem. For sparse methods, the regularization parameter is selected using cross-validation. Systematically, the "optimal" model error (loss) is computed to be smaller than that of LORETA [60].

Laplacian graph regularized discriminative source reconstruction is another new method that tacitly codes the label information into the graph regularization term to extract the discriminative brain sources. This model is capable of developing with different assumptions. The weakness of this model is that only one spot as a common activated source is used. However, there actually may be several common source activation regions [61].

In other studies, multivariate autoregressive models are fitted to electroencephalographic time series. This technique directly provides a dynamical model of current distribution from the data. The proposed method considers a realistic estimated model of data compared to previous methods which consider approximate models of internal connectivity of sources. The results show that estimating multivariate autoregressive (MVAR) models improves the quality of inverse solutions to a significant degree compared to immediate conventional solutions. However, these conditions are true when the regularized inverse of Tikhonov is used [62].

Recently, variation-based sparse cortical current density (VB-SCCD) algorithm, which extracts the sparsity of the variational map of the sources, has been considered as a promising approach in comparison to source imaging techniques. Furthermore, the application of the alternating direction method of multipliers (ADMM) algorithm is demonstrated, which presents the useful solution of optimization problem [35].

Other methods have been proposed to improve the accuracy and spatial resolution, which try to accurately determine the number of source dipoles and reduce the computational cost, that are discussed in the following.

For closely-spaced neural sources, first principle vectors (FINE) can be used to improve the spatial resolution and localization accuracy from EEG and MEG measurements. Using this method, the performance of localizing multiple closely spaced, and inter-correlated sources is enhanced in low SNR scenarios [42].

In order to discover the low dimensional manifolds from recorded EEG handle, the isometric feature mapping (ISOMAP) algorithm has been used to solve large-scale high dimensional problems efficiently and quickly. Then, multidimensional support vector regression (MSVR) and iterative re-weight least square (IRWLS) are applied to find the relationship between the observation potentials on the scalp and the internal sources within the brain based on reduced data dimension. It has been demonstrated that this algorithm can obtain more robust estimations for EEG source localization problem. Any particular assumption from prior knowledge is not employed in MSVR leading to more flexible method [63, 64].

Multi-planar analytic sensing is used to increase the localizing accuracy in the solution of EEG source localization. The estimation of the projection on each plane of the dipole positions is a non-linear problem. This problem is modified by the proposed method to find the polynomial roots. However, the computational cost of this method is very high, which reduces the possibility of online implementation [65].

In 2008, one technique was proposed which localizes source activity using a linear mixture of temporal basis functions (TBFs) learned from data. Performance of this method demonstrates significant improvement over existing source localization methods [66]. The extended Kalman filter as well as particle filter solutions are also applied for active source finding using a dynamic probabilistic model. These algorithms take into account the neural dynamics and this issue can be helpful to obtain more accurate localizing [67].

Usually, source directions are selected to maximize power in the analysis of rhythmic brain activity. But, [68] proposes to maximize bicoherence instead of power. Simulation results demonstrate considerable bicoherence differences in motor areas. This differences could not be discovered from analyzing power differences.

The number of dipoles is significant in the dipole source localization (DSL) method. Akaike's information criterion (AIC) and Bayesian information criterion (BIC) can adaptively estimate the dipole number. Truncated RAP-MUSIC (TRAP-MUSIC) is another technique for dipole number determining. According to the results of various studies, MUSIC-type localization such as TRAP-MUSIC method could become more valid and appropriate for various online and offline EEG applications [69, 70]. Also, the Powell algorithm can be applied to estimate the dipole number from the scalp EEG using different penalty functions of information criterion.

On the other hand, optimization techniques have been effective in improving inverse-problem solutions. Simulated annealing (SA), particle swarm optimization (PSO), genetic algorithm (GA) and differential evolution (DE) are another statistical methods which are applied in localizing EEG dipole sources, particularly in one-dipole estimation. Regarding two-dipole localization, GA and DE have better performance than two other methods, but DE needs the setting of a suitable parameter. By reducing the signal to noise ratio (SNR), the efficiency of all algorithms is decreased, while SA and PSO appear to be very sensitive to the correlation between the sources. The results show that the correlation between the sources strongly affects SA and PSO outputs. Generally, among these four methods, GA has more appropriate computational cost and performance [71].

*Improving EEG source localization methods by non-statistical strategies*

Cavities of the human head impress EEG dipole localization. Computer simulation has been used to study these effects. According to the obtained results, these effects are negligible for the dipoles that are located in the cortex or the subcortex. While for the dipoles of the brain stem, the cavity and electrode arrangement on the scalp extremely affect the EEG inverse dipole solution [72].

In another work, the effects of electrode location errors on EEG dipole source localization is discussed using a practical head model. It is also shown that the white noise is more effective than electrode misplacements in localization errors. Furthermore, it is reports that increasing the number of electrodes improve the source localization results, but the absolute enhancement is less considerable for larger electrode numbers [73].

Enough sampling of the potential surface field, a careful conducting volume estimation (head model) and a convenient and well-understood inverse technique are effective factors in the accuracy of EEG source localization. Furthermore, increasing the density of the sensors enhances the accuracy of source localization. In addition, adding samples on the inferior surface increases the accuracy of defined sources at all depths [74]. These non-statistical strategies are summarized in Fig.5.

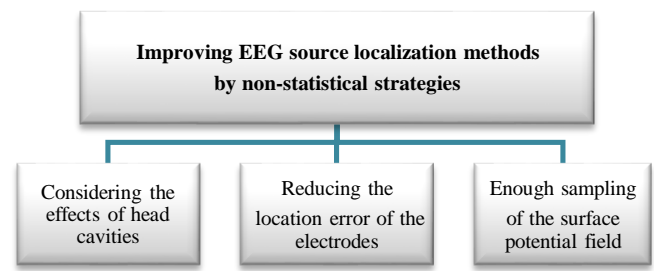

Fig.5. Non-statistical strategies for Improving EEG source localization methods.

*Investigating the effects of the head model on EEG source imaging results*

As already mentioned, the accuracy of the EEG source localization methods strongly depends on the accuracy of the head model [75]. Spherical models [76], realistically boundary element method [77], finite element models [78], and finite difference method (FDM) [79] are the most popular head models.

Considering the sources of the temporal lobe, the realistically shaped head model localizes fewer dipoles than 3-sphere model [80, 81].

The boundary element method (BEM) uses closed triangle meshes and a limited number of nodes to approximate the different compartments of volume conductor models. For all BEM models, fitted dipole locations are assessed to indicate the dependency of the averaged and maximum localization errors on their node distributions [82].

The effects of white matter (WM) anisotropic conductivity on EEG source localization also has been studied in [83]. It has been shown that the accuracy of EEG dipole localization in the primary visual cortex does not significantly improve using experimental data obtained using visual stimulation and anisotropic models incorporating realistic WM anisotropic conductivity distributions.

The performance of a more complex head model is higher than a model with less tissue surfaces in inverse source localizations. Another important component of the head that changes the scalp potentials as well as the results of inverse source localizations is the cerebrospinal fluid (CSF) [84]. When a particular bioelectric conductivity model is used for each patient, the electroencephalographic source localization (ESL) results will be accurate enough. Although the influence of anisotropic conductivities in the skull and WM are studied, accurate modeling of the highly conductive CSF region has not been investigated yet, except [85], which has studied the influence of partial volume errors in CSF segmentations on the ESL bioelectric model. Some voxels include both CSF and gray matter. Thus, they cannot be specified with an absolute single label. This problem increases the volume errors in CSF segmentation. In this approach, a layered gray matter-CSF model is used to form equivalent anisotropic conductivity tensors in regions where partial volume errors are expected [85].

Skull has higher geometric complexity and lower conductivity than the other tissues inside head. According to the results reported in [86], skull geometry simplifications have a more extensive effect on ESL compared to the conductivity modeling. Miscalculation of skull conductivity can generate source localization errors as large as $3cm$ [87]. Among the different methods of brain sources localization, it has been proven that minimum-norm cortical source estimation in layered head models is robust against skull conductivity error [88].

Except for the anatomical information, the major tissue compartments can be calculated in the BEMs. Finite element models (FEM) can consider more tissue types and complex anatomical structures. While for the higher precision, semi-automated segmentation and a higher computational cost are necessary. Furthermore, a highly detailed FEM has been proposed, which is denominated ICBM-NY or "New York Head". In this model, ICBM152 anatomical template (a non-linear average of the MRI of 152 adult human brains) is determined in MNI coordinates. The field of view has been extended to the neck and the detailed segmentation of six tissue types are performed (scalp, skull, CSF, gray matter, white matter and air cavities) at $0.5mm^3$ resolution (Fig.6) [89].

*Diagnosis of brain abnormalities using common EEG source localization methods*

It is shown that different brain source localization techniques are effective in the diagnosis and treatment of several brain abnormalities and diseases. Among the brain abnormalities that have been investigated using EEG source localization methods, epilepsy and attention-deficit hyperactivity disorder (ADHD) have more contribution. Therefore, in the following, brain abnormalities are studied in three sub-sections. First, the methods that are used to detect epilepsy spikes are reviewed. Then, ADHD is studied and finally, the methods that are used to diagnose and treat other brain abnormalities are presented

- *Detection of epileptic seizures using brain activity localization based on EEG signals*

Usually, subdural electrode technique is used to reliable localization of epileptogenic tissue. Since this method is invasive, it may cause infections. Hence, a non-invasive approach is needed for deep source localization. In 1982, for the first time the variations in the beta activity of intracerebral EEG recordings following diazepam intravenous injection are evaluated using a localization method [25]. In this technique, it has been investigated whether beta activity in the damaged areas of the brain is increased or not? Table I provides a series of studies on this topic from 1995 to 2019, respectively. Publication year, study objectives, methodology and results and conclusions of each study have been investigated (Fig.7).

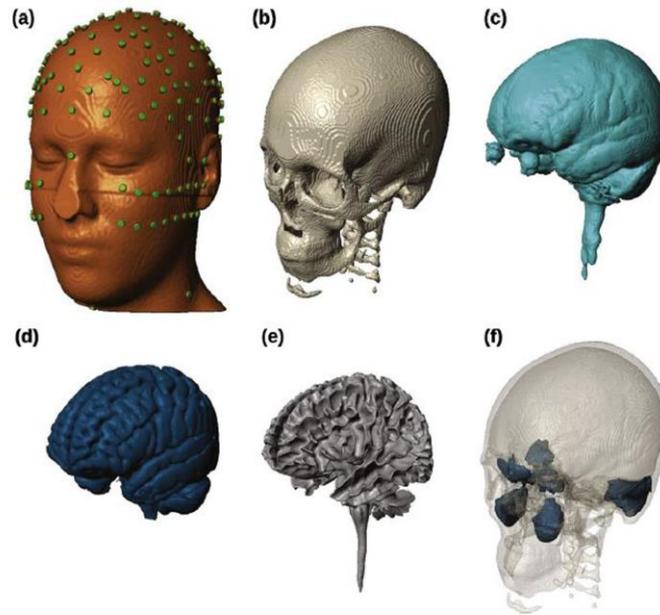

Fig. 6. Segmentation of the ICBM-NY head into six different tissue types. From (a) to (f): scalp (with 231 electrodes placed), skull, cerebro-spinal fluid, gray matter, white matter, air cavities. Note that the disc electrodes and underlying gel in (a) are not physically modeled. Instead, they are represented by a single tetrahedral mesh-element on the scalp surface [89].

• *Diagnosis and treatment of ADHD abnormalities*
ADHD is considered as a clinical psychiatric disorder which affects frontal circuitry regarding deficits in practical cognitive functions. Some imaging studies suggested that the patients with ADHD have smaller anterior cingulate cortex and dorsolateral prefrontal cortex which is related to memory [90]. Also other studies showed a delay in cortical thickness in patients with ADHD [91, 92]. Furthermore, a study revealed that the ADHD patients have a weakened activity in the frontostriatal region in their brain which is important in inhibitory control and attention [93]. It causes problems such as high financial costs, stress to families and interpersonal relationships, and unfavorable academic and vocational consequences [94]. These disorders are specified by altered levels of inattention, hyperactivity and impulsivity indications. ADHD is a childhood disorder that does not continue into adulthood [95]. Brain source localization of the EEG signal of ADHD patients has created a new way for diagnosing and treatment of the disease. Some studies suggested the application of the quantitate EEG in the assessment of diversities in baseline spectral power profiles and pharmacological and non-pharmacological treatments which had effects on electrocortical activity [96]. Table II summarizes major studies and research concerning ADHD source localization.

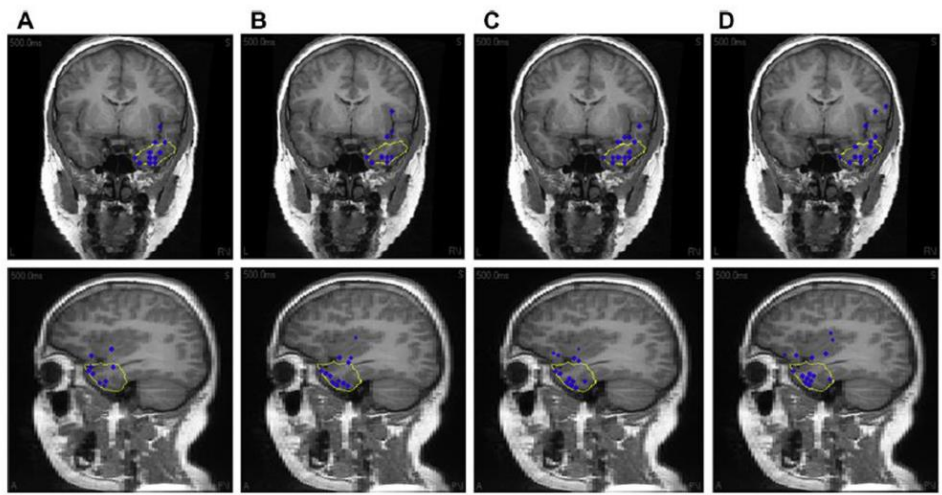

Fig. 7. The source location of all interictal spikes in patient 3 obtained using respectively (A) 128, (B) 96, (C) 64 and (D) 32 electrodes. The blue dot shows the location of the maximum of sLORETA. The yellow line represents the resection boundaries [73].

TABLE I

Epilepsy included studies.

| Publication(s) | Year | Study objectives | Methodology | Results and conclusions |
|---|---|---|---|---|
| Tseng et al. [97] | 1995 | Localization of discharge sources of epileptic spikes | Electric dipole model | After 50 iterations, the best value of fit is 100%. |
| Lantz et al.[98] | 1999 | Ictal epileptiform activity localization in patients with complex partial epilepsy of temporal lobe origin in the frequency domain | • FFT dipole approximation<br>• Each of the frequency point potential maps are used for source localization algorithms.<br>• Electric dipole model is applied for source localization. | In different patients, ictal frequencies have a range between 3.5 to 8.5Hz. A changeable degree of stability over time is shown by these frequencies. During the various phases of seizure progress, the dipole results of a specific frequency are similar. The dipole results of different frequencies are only similar in patients with more than one outstanding frequency. |
| Lantz et al.[99] | 2001 | Space-oriented segmentation and 3-dimensional source reconstruction of ictal EEG patterns | Weighted minimum norm (WMN) | An effective method for non-invasive determination of the starting and perhaps also the extension of epileptiform activity in patients with epileptic seizures, is segmentation of ictal EEG together with further 3-dimensional source reconstruction. |
| Huppertz et al. [100] | 2001 | Localization of epileptiform EEG activity related to focal epileptogenic brain damages and interictal delta | Electric dipole model | Maximum dipole concentration is closer than $10mm$ to the nearest damaged margin in 66% of the patients with focal delta activity. This maximum is often at the border or within pathologically changed cortical tissue. |
| Iwasa et al.[101] | 2002 | Different patterns of gelastic seizure with or without a sense of mirth is localized using dipole sources. | Electric dipole model | The production of gelastic seizures with a sense of mirth leads to the neural activities in hippocampal regions and these activities can be related to the motor act of laughter in the cingulate. |
| Alper et al.[102] | 2008 | Three-dimensional statistical parametric maps of background EEG source spectra are used to localize epileptogenic areas in the partial epilepsy. | LORETA | In this study, sources which are anatomically near the location of epileptogenic areas are localized using three-dimensional tomography. In this case, intracranial recordings are used. |
| Clemens et al. [103] | 2008 | The image of the cortical effect of lamotrigine in patients with idiopathic generalized epilepsy is recorded. | LORETA | In a large group of voxels including parts of the temporal, parietal, bilateral occipital cortex and in the transverse temporal gyri, insula, hippocampus, and uncus on the right side, theta activity is reduced. Also, in a rather smaller cortical area including the right temporo-parietal connection and enclosing parts of the cortex and part of the insula on the right side, alpha activity is reduced. |
| Jarchi et al.[104] | 2009 | A hybrid second-order blind identification and extended rival penalized competitive learning algorithm is handled for seizure source localization. | Second order blind identification (SOBI) is generally exploited to calculate the brain source signals in every window of the EEG signals. The rows of the estimated unmixing matrices in all of the windows are clustered | This method seems to be very useful for finding seizure foci. However, in order to obtain acceptable results, it is necessary to use a large number of scalp signals and assess its concurrently recorded intracranial signals. |
| | | | using a new clustering technique based on rival penalized competitive learning (RPCL). This signal is projected back to the electrode space and is converted to the dipole source localization using a single dipole model. | |
| **Oliva et al.[105]** | 2010 | EEG dipole source localization of interictal spikes in non-lesional Temporal lobe epilepsy (TLE) with and without hippocampal sclerosis. | MUSIC | The results of the study show considerable inter-individual variability in the localization of the dipoles, with the majority of patients in both groups which have a localization in the temporal lobe but only a small proportion and in the mesial temporal region. |
| **Clemens et al. [106]** | 2010 | The sources of theta rhythm is localized in partial epilepsy patients with and without medication. | LORETA | Bilateral theta maxima are observed in the temporal theta area (TTA), parietal theta area (PTA) and frontal theta area (FTA) in one group. In another group, this activity is augmented all over the scalp subject to group 1. The maxima of theta activity happen in the TTA, PTA, and FTA. Although, the abnormality centers move towards the medial cortex in the PTA and FTA. |
| **Koessler et al. [107]** | 2010 | The sources of ictal epileptic activity are determined by high-resolution EEG and proved by Stereo-Electro-Encephalo-Graphy (SEEG) | moving dipole, rotating dipole, MUSIC, LORETA, and sLORETA | • Agreement of Tc-ethylene cysteine diethylester (ECD): 9/10 • Agreement of MUSIC and LORETA: 7/10 • Agreement of sLORETA: 5/10 |
| **Besenyei et al. [108]** | 2011 | EEG background activity of childhood with benign rolandic epilepsy is unusual in the temporal and inferior parietal cortex | LORETA | Abnormal activity is clearly observed in the temporal and parietal cortical areas. These areas are associated with major components of the Mirsky attention model and also the Perisylvian speech network. Hence, attention and speech may be damaged in benign rolandic epilepsy patients. |
| **Wennberg et al. [109]** | 2011 | The extracranial EEG of temporal lobe spike foci is compared with their intracranial sources in the patients of mesial temporal lobe epilepsy. | Electric dipole model | EEG or MEG cannot find intracranial mesial temporal spikes. Mid temporal EEG spikes are localized to the lateral temporal neocortex. |
| **Blenkmann et al. [110]** | 2012 | Focal cortical dysplasia (FCD) localization in epilepsy patients using equivalent current dipole method. | Electric dipole model | Epileptogenic zone related to the location of ECDs dilates beyond the FCD visible in MRI. According to the results, the ECDs place in a shell parallel to the part of the FCD surface. |
| **Clemens et al. [111]** | 2012 | The endophenotypes of the common idiopathic generalized epilepsy syndromes are localized using EEG and LORETA. | LORETA | Juvenile myoclonic epilepsy results in augmented theta activity in the posterior parts of the cortex. Also, the endophenotype for absence seizures augments theta activity in the fronto-temporal limbic areas. Diffused biochemical abnormality cannot be observed in juvenile myoclonic epilepsy and absence seizures. |

| Publication(s) | year | Study details | Methodology | Results and conclusions |
|---|---|---|---|---|
| E. Coutin-Churchman et al. [112] | 2012 | In patients with surgically removed epileptogenic foci, the sources of EEG interictal spike activity is found and quantified. | sLORETA | During the onset, fourteen patients had 90–100% of spikes within the site of resection (SR), and 9 had 50–89%. Most patients with more than 50% of activity sources within SR were seizure free, but the five patients who had all activity sources outside SR were not seizure free. |
| Lu et al.[113] | 2012 | Seizure source localization of partial epilepsy patients using FINE spatio-temporal dipole localization and directed transfer function. | FINE | Compared to the surgically resected brain regions, the source detection accuracy of seizure onset zone using 76-channel EEG is higher than other EEG arrangement with fewer electrodes. |
| Shirvany et al. [114] | 2012 | Particle swarm optimization for finding the precise location of the epileptogenic foci. | Standard Particle Swarm Optimization | The global minima are computed with appropriate accuracy and a convenient number of iterations. |
| Itabashi et al.[115] | 2014 | The spike source of small focal cortical dysplasia is estimated in the dorsal perirolandic area. | Electric dipole model | Six patients are selected. Their clinical characteristics were leg sensori-motor seizures in 5 patients and eye version in 1 patient. A small focal cortical dysplasia in the dorsal perirolandic region is localized. |
| Sohrabpour et al. (Fig.6) [73] | 2015 | The influence of EEG electrode number on epileptic source localization is investigated in pediatric patients. | sLORETA | When the number of electrodes increases, the results of source localization is enhanced, while the absolute improvement of the accuracy is degraded (Fig.7). |
| Chowdhury et al. [116] | 2016 | Examining the complex patterns of epileptic activity generators using ExSo-MUSIC | ExSo-MUSIC | ExSo- MUSIC favored the strongest source activity. ExSo-MUSIC is better for single and deep sources with large signal-to-noise ratio. |
| Strobbe et al. [117] | 2016 | Presurgical epileptogenic focus localization using multiple sparse volumetric priors. | Bayesian model selection | The results show that the proposed approach is very effective to specify the irritative zone rather than other distribution methods such as LORETA and ECD model. |
| Eom et al.[118] | 2017 | Source localization of centrotemporal spikes in interictal spikes of benign childhood epilepsy | sLORETA | Current-source density (CSD) of the maximal negative peak is measured. In all of the patients, the rolandic area is seen in the distribution of the CSD. |

TABLE II
ADHD source localization.

| Publication(s) | year | Study details | Methodology | Results and conclusions |
|---|---|---|---|---|
| Jonkman et al. [119] | 2004 | early visual selective attention deficit in ADHD children | Electric dipole model | Smaller frontal positive activity (frontal selection positivity; FSP) in ADHD children is around 200$ms$, while later occipital and fronto-central negative activities is observed to be uninfluenced. In control subjects, the FSP is showed by posterior-medial equivalent dipoles. It may represent the contribution of multiple enclosing areas. |

| Nazari et al. [120] | 2010 | During the cued continuous performance test, visual sensory processing deficit in ADHD children is localized in the occipital region. | SwLORETA | P100 and N200 ERP investigation in response to both Go, and NoGo stimuli, specifies that there is a low rate of Go correct response and high rate of omission errors in ADHD children. Also, delayed P100 and N200 latency, and lower P100-NoGo amplitude is observed. Moreover, P100 latency of Go against NoGo tasks is delayed. P100 source is located in the occipital area. Particularly in the NoGo condition, early electrical activity in ADHD has a significant decrease. |
|---|---|---|---|---|
| Helfrich et al. [137] | 2012 | Cortical irritability is monitored during repetitive transcranial magnetic stimulation in ADHD Children. | RAP-MUSIC algorithm | During 1 Hz-repetitive transcranial magnetic stimulation (rTMS), TMS-evoked N100 amplitude is reduced. Almost, after 500 pulses, this ERP attained a fixed plateau at the group level. According to brain source analysis, TMS-evoked N100 variation corresponds to rTMS effects of the stimulated motor cortex. |
| Bluschke et al. [121, 122] | 2016 | The neuronal mechanisms are studied after theta/beta neurofeedback of ADHD children. | sLORETA | After neurofeedback treatment, impulsive behavior is decreased. Also, impulsivity of neuronal mechanisms are modulated in ADHD. |
| Janssen et al. [122] | 2016 | The oddball task source study of ADHD children. | LORETA | Problems associated with task-relevant events in ADHD children systematically correspond to reductions in the amplitude of the P3b ERP. Dissimilarities are mostly located in frontal polar and temporoparietal regions in the left hemisphere. |
| Czobor et al. [123, 124] | 2016 | Aberrant error-processing in ADHD adults is discussed. | LORETA | The ERP-attenuation is outstanding not only at usual ROI-electrodes but also across many other brain regions, with a specified subset of group dissimilarities and indication-correlations revealed at temporo-parietal sites, with right-lateralization. |
| Khoshnoud et al. [124] | 2017 | Different cortical source activation patterns are investigated in ADHD children during a time reproduction task. | Electric dipole model | During the time reproduction phase, considerable differences are inaugurated in the mean alpha power for the prefrontal source group. Hence, electrophysiological evidence is presented for time perception deficiencies, selective visual processing disturbances and working memory impairment in the ADHD children. |
| Chmielewski et al. [125] | 2018 | Effects of multisensory stimuli on inhibitory control are studied in adolescent ADHD. | sLORETA | These effects are related to modulations at the response selection stage (P3 ERP) in the medial frontal gyrus (BA32). |
| Bluschke et al. (Fig.8) [126] | 2018 | Separating inattentive and combined ADHD subtypes. | sLORETA | Altered pacemaker-accumulation processes in medial frontal structures have differentiated the attention deficit disorder (ADD) from the attention deficit hyperactivity disorder-combined (ADHD-C) subtype. |
| Bluschke et al. [127] | 2018 | Neural mechanisms of successful and deficient multi-component behavior are studied in early adolescent ADHD. | sLORETA | In the uni-modal and bi-modal experiments, response selection mechanisms are observed in the inferior parietal cortex (BA40) by neurophysiological processes. |

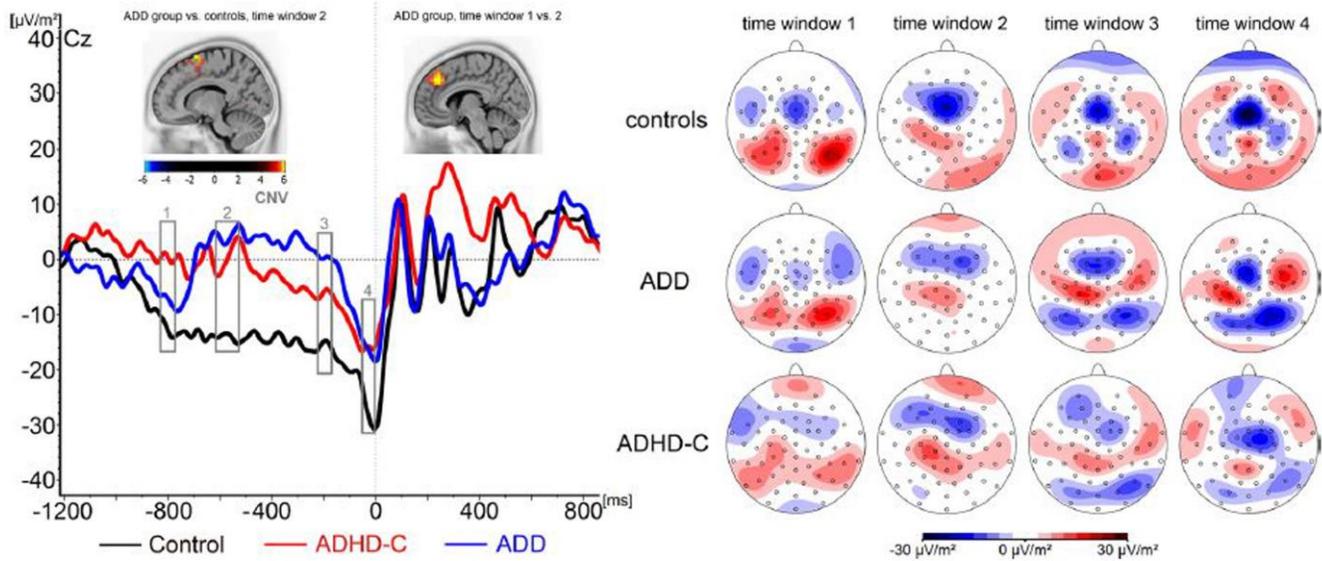

Fig. 8. The difference in the activity level of the brain areas in two different types of ADHD[126]

- *Diagnosis of other brain abnormalities using EEG source localization methods*

In [128], multichannel EEG of an advanced meditator was recorded during four distinct and frequent meditations and LORETA was used to locate intracerebral source gravity centers. The 'gamma' band of EEG functional images (35-44 Hz) is displayed. It is seen that the activity of the frequency band had clear variation between meditations.

According to previous reports, conduct problems of adolescents are associated with P300 ERP signal. In another study, P300 sources have been modeled using implementing current source density-boundary element techniques in "at-risk" adolescents. The results of the techniques accurately indicate that conduct problems are related to a particular dysfunction of the frontal brain [129].

To locate equivalent current sources of visual attention in the hemi-space, event-related potential P2 of EEG and LORETA method is used. It is reported that the amplitude of P2 increases when subjects understand the stimuli or pay attention to the stimuli [130].

Subclinical rhythmic electrographic discharges of adults (SREDA) is presently presumed a benign EEG pattern of unknown importance. The underlying cortical sources and their generating mechanisms are uncertain. In order to better understand this uncommon EEG pattern, zumsteg et al administrated a source localization analysis of SREDA. Hyperventilation sensitivity and a posterior hemispheric source localization maximal in the parietal cortex bilaterally, in large part overlying the anatomical distribution of the vascular watershed areas, is seen in the patient with typical SREDA [131].

By localizing the retinotopic organization in the human primary visual cortex (V1), the spatial resolution of EEG brain source imaging is studied. Between the fMRI-determined activation centers in V1 and the EEG source imaging activation peak, estimated at equivalent C1 components (peak latency: $74.8\pm10.6ms$), there is $7mm$ mean location error (Fig.9) [132]. Type 1 Schizencephaly (SZ) is a cerebral malformation specified by a gap lined and enclosed by a polymicrogyric cortex, extending from the pial region to the peri-ventricular heterotopia. A method has been proposed to incorporate and compare dipole source imaging method and SEEG technique in order to define the irritative and epileptogenic zones in a case of type 1 SZ. The results demonstrated that in these cases, source localization methods can help to specify the irritative zone and relevant targets for SEEG [133].

Sleep EEG quasi-rhythmic activity increases dramatically within the frequency band of 11–16 Hz during sleep spindles, specified by gently increasing, then slowly decreasing amplitude. Ventouras et al. applied ICA to process sleep spindles, in order to study the possibility of extracting through visual analysis of the spindle EEG and visual selection of independent components, spindle "components" (SCs) corresponding to separate EEG activity patterns during a spindle and to determine the intracranial current sources underlying these SCs using LORETA. Based on temporal and spectral analysis of ICs, SCs can be extracted by reconstructing the EEG through back-projection of separate groups of ICs. The intracranial brain sources of the SCs were formed to be spatially stable during the time evolution of the sleep spindles [134].

The sources of symptom provocation on spider phobia have been investigated through late event-related potentials (ERPs) using sLORETA. Mean amplitudes of ERPs are extracted in the time windows of $340$-$500ms$ (P300) and $550$-$770ms$ (late positive potential, LPP). In response to spider pictures, P300 and LPP amplitudes of phobics are higher than controls. Generally, sources were located in the occipital and parietal regions; the ventral visual pathway of the brain was related to visuoattentional processing. Furthermore, some sources were in the cingulate cortex, insula areas which are related to emotional processing and the demonstrations of aversive bodily states. Furthermore, the priming of flight behavior sources were marked in premotor areas. The results indicated that source localization is an appropriate alternative for recognizing brain regions of phobia [135].

EEG source localization of N20–P20 somatosensory evoked potentials (SEPs) can obtained useful information about the sources of the primary sensory hand area. For three healthy subjects, median nerve stimulation was enforced and single-dipole localization was implemented for the N20–P20 SEPs. Lower computational time and higher accuracy are the most significant features of this method compared to another methods [136].

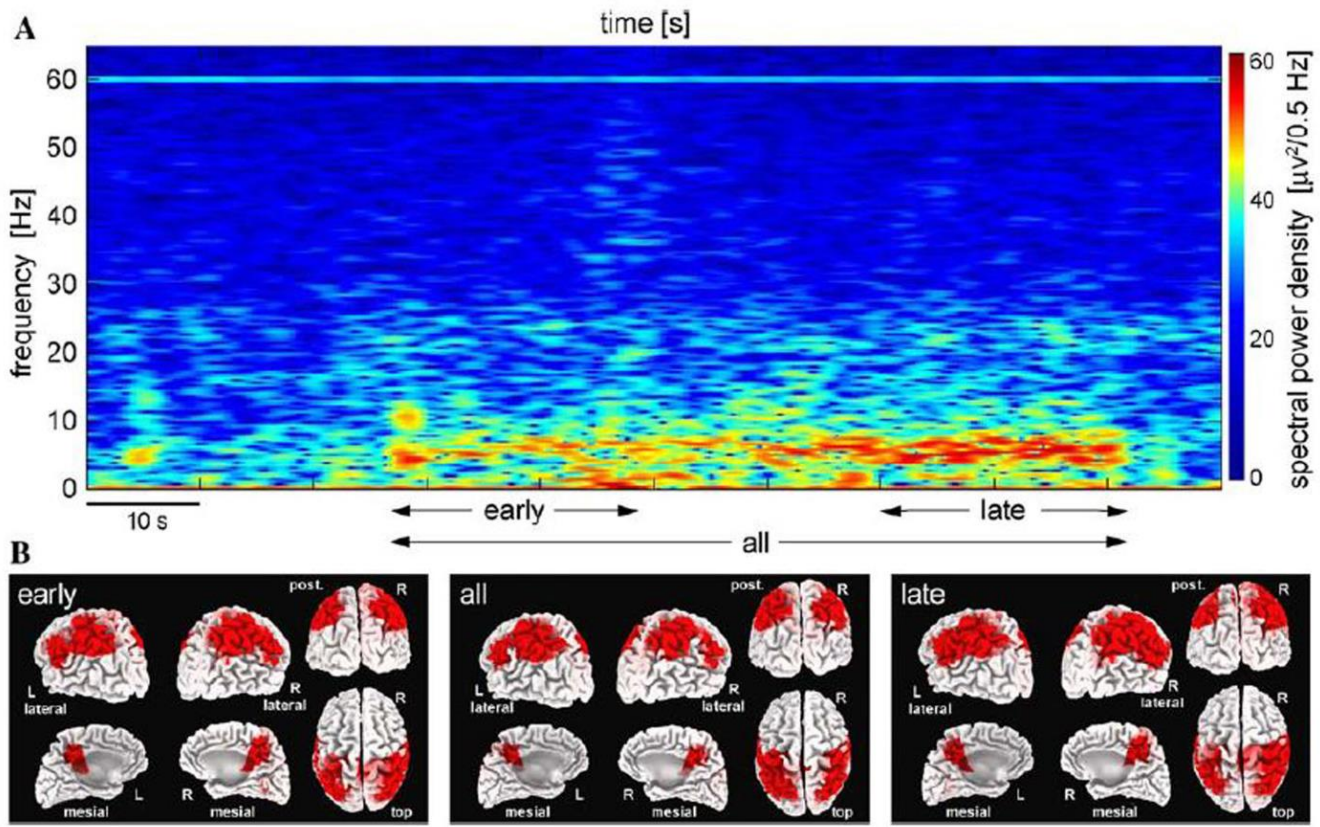

Fig. 9. (A) Spectrogram of a single representative SREDA episode showing an irregular increase of spectral density power in lower delta and theta EEG frequency bands during the course of the pattern. Note that, with respect to frequency, there is no clear evolution of the spectral pattern over the course of the SREDA event. (B) Three-dimensional frequency domain (theta cross spectra) SNPM LORETA reconstructions for three different periods of 10 SREDA Note that there is no significant difference of cortical activation patterns for the three periods analyzed [132].

Based on cognitive hypotheses about the phenomena of hypnosis, in order to generate a selective alteration or disconnection of some mental operations, it is possible that executive attentional systems either inhibited or overactivated. Recently, during hypnotically induced paralysis brain imaging studies, changes in the activity of both medial (anterior cingulate) and lateral (inferior) prefrontal areas are reported using brain imaging, that it is overlapped with attentional control areas and inhibitory processes. The minimum-norm methods are used to investigate topographical EEG analysis, the spatial organization and the temporal sequence of neural processes, and to localize the principal anatomical brain generators [137].

A study is performed to answer the question of whether psychotic symptoms affect electroencephalographic activity in particular brain areas especially the delta band activity? Current source density images of the delta, theta, alpha and beta activities have been generated using LORETA. The left inferior temporal gyrus, right middle frontal gyrus, right superior frontal gyrus, right inferior frontal gyrus and right parahippocampal gyrus are the areas that show more activity in the delta frequency band in patients using the LORETA algorithm [138]. In order to increase working memory in healthy people and improve mood in major depression, prefrontal transcranial direct current stimulation (tDCS) with the anode placed on the left dorsolateral prefrontal cortex (DLPFC) is employed. After anodal tDCS of the left DLPFC and cathodal tDCS of the right supraorbital area, the distribution of neuronal electrical activity changes were assessed using spectral power analysis and sLORETA. According to the obtained results, in addition to enhancing working memory performance, anodal tDCS of the left DLPFC and/or cathodal tDCS of the contralateral supraorbital region may modify sectional electrical activity in the prefrontal and anterior cingulate cortex [139].

Furthermore, in patients with Obsessive-Compulsive Disorder (OCD), sLORETA is used to assess the activity of intracortical EEG sources. Low-frequency power excess (2–6$Hz$) in the medial frontal cortex is observed in OCD, while increased low-frequency power in a component in group ICA shows the activity of subgenual anterior cingulate, adjacent limbic structures and to a lesser extent also of lateral frontal cortex (Fig.10) [140].

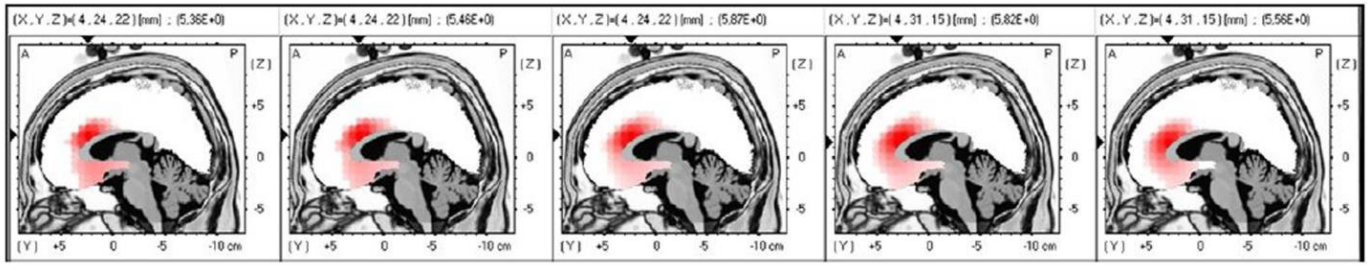

Fig. 10. Sagittal view (left of the picture is front of the head) of absolute current density increase in OCD patients compared with controls at 2–6 Hz. Each figure is sliced to its own t-value maximum (tmax2Hz = 5.36, tmax3Hz = 5.46, tmax4Hz = 5.87, tmax5Hz = 5.82, tmax6Hz = 5.56). Figures show only the most significant voxels (the darker the red color, the higher the t-value) [140].

During tonic cold pain, source activity in low-frequency bands (<12$Hz$) is obviously decreased, while activity in high-frequency bands (>12$Hz$) usually increases in several brain areas. A frequency-domain EEG source analysis has been performed in [141], in order to investigate electrocortical responses to tonic cold pain and recognize potential electrocortical symptoms of critical tonic pain. EEG power spectra are investigated in five frequency bands: 1–4$Hz$, 4–8$Hz$, 8–12$Hz$, 12–18$Hz$ and 18–30$Hz$, to localize EEG cortical sources using sLORETA. It is reported that the EEG spectral power in 8–12$Hz$ of healthy subjects under tonic cold pain is considerably lower than no-pain control subjects, while it is higher in 18–30$Hz$ band, in large brain regions. The left medial frontal, left superior frontal, posterior cingulate and the anterior cingulate activities are related to 4–8$Hz$, 8–12$Hz$ and 12–18$Hz$ frequency bands. These results also indicated considerable negative correlations with subjective pain ratings.

In [142], cortical source localization is studied in the response inhibition of individuals with psychopathic traits using sLORETA algorithm. In this research, EEG signal of NoGo stimuli in a Go/NoGo task is analyzed. The simulations of P3 elicited by the NoGo stimuli illustrate that at the frontocentral area, NoGo-P3 amplitudes of psychopathic trait group decrease extremely more than the control group.

P300 ERP investigation helps to better understand the mechanisms of attention and memory operations. Conventional averaging techniques are used to analyze the P300 data, and sLORETA localization algorithm is performed for P300 source localization. According to statistical analysis, it is impossible to localize the P300 component because these components are related to a wide cerebral cortex network [143]. Furthermore, sLORETA algorithm results show that in patients suffering from normoacousic tinnitus, EEG sources reduce in left temporal and inferior parietal gyri [144].

Remifentanil is a potent, short-acting synthetic opioid analgesic drug. It is given to patients during surgery to mitigate pain and as an adjunct to an anesthetic. The effect of remifentanil on resting EEG source location is studied in [145]. These effects can be used as a biomarker of remifentanil. Remifentanil infusion clearly changes EEG brain source locations compared to baseline data, which are consistent and robust. According to the sLORETA results, remifentanil derived variations are often outstanding in cortical activity at frontal, fronto-central and fronto-temporal brain regions on the left hemisphere.

Three ERP components, i.e., the mismatch negativity (MMN), the P300 and the N400 has been applied to localize the neuronal generators during an ERP study, using LORETA. In the case of the P300, with aging, the maximum intensities are shifted from frontal to temporal lobe, while there is no change for the MMN component. The age has not any effect on N400 characteristics. But, gender has a considerable effect on the response time of the subjects. The response of males is faster than females [146].

In Diabetes mellitus (DM), structural and functional changes occur in the central nervous system. DM is a metabolic disorder which involves central and peripheral nervous system. Poor control of glucose leads to a reduction in synaptic connections and a neurodegenerative disorder may occur which results in brain atrophy and dementia [147, 148]. In one study, state cortical and its relevance to clinical features has been evaluated using resting EEG activity. Wavelet analysis has been used to summarize the frequency bands with the corresponding topographic mapping of power distribution. sLORETA is applied as a source localization method of EEG signal. In this case, localization results disclose that the reason for these changes is the frontal region activity of the delta band and the activity of central cortical areas of the gamma band. Also, source activity is decreased in the left postcentral gyrus for the gamma band and in the right superior parietal lobule for the alpha1 (8–10$Hz$) band (Fig.11) [149].

The key diagnostic feature/symptom of the restless leg syndrome (RLS) is the circadian change of sensory and motor symptoms with increasing intensity in the evening and at night. However, there is a strong relationship between motor and cognitive symptoms in many neurological diseases. But how the cognitive function of RLS patients change overnight and the neurophysiologic mechanisms related to circadian changes has not been examined yet. In [150], the analysis of flanker interference effects and sLORETA technique are used to investigate daytime effects (morning vs. evening) on cognitive performance in RLS patients.

In a study, EEG has been recorded from patients with ischemic brain lesions during a tonic hand muscle contraction task and during continuous visual stimulation with an alternating checkerboard. Cortico-muscular coherence was correctly localized to the primary hand motor area and the steady-state visual evoked potentials to the primary visual cortex in all subjects and patients. Sophisticated head models tended to yield better localization accuracy than a single sphere model. It is reported that a minimum variance beamformer (MVBF) provided more accurate and focal localizations of simulated point sources than an L2 minimum norm (MN) inverse solution. In the real datasets, the MN maps had less localization error but were less focal than MVBF map. In RLS patients, the flanker interference effects are larger in the evening than in the morning, whereas there is no circadian variation in healthy controls. Also, in the interfering task condition in the evening, N1 amplitudes of neurophysiological data in the RLS patients is smaller than controls. This does not hold for the morning time [151]. Table III shows brain abnormalities and their common EEG source localization methods.

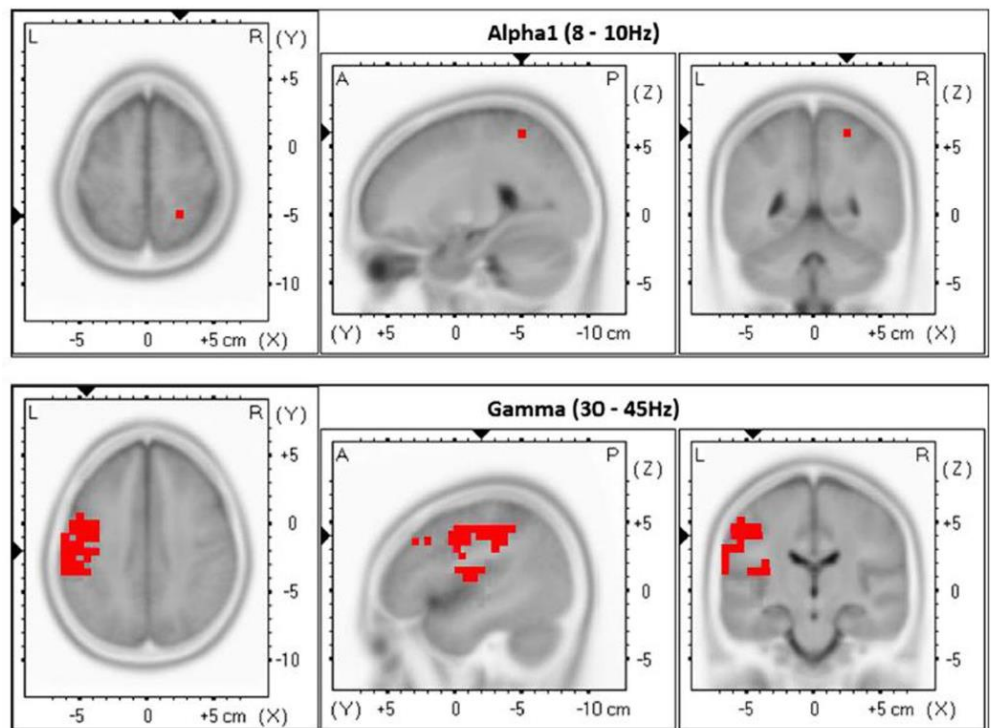

Fig. 11. Source localization analysis of the individual frequency bands with sLORETA. Axial (from top), sagittal (from left) and coronal (from back) views of source locations in the alpha1 (8–10 Hz) and gamma (30–45 Hz) frequency bands where diabetes mellitus patients had significantly decreased activity compared to healthy controls. These regions are identified with red color [149].

*Beneficial and challenging effects*

But one question remains that "What are the implications for future research on brain source localization techniques in brain functional abnormalities?" From 1995 to 2018, the issue of EEG source localization of interictal spikes has been very much considered. Neural mechanisms of ADHD were studied using source localization methods. sLORETA algorithm has been used more than other localization methods, while new emerging localization methods with better performance and accuracy can be used instead, in order to make the algorithms more reliable. Different ADHD pattern classification by determining the active regions of the brain in a specific task is an important issue, but enough research has not been performed regarding this issue in the recent years. Applying a specific treatment approach depending on the type of ADHD would increase the treatment speed and reduce its cost.

The effect of psychiatric drugs on the activity of brain sources has been less considered, hoping to be studied further in the future research using EEG source localization methods. In the study of various diseases using brain source localization techniques, sLORETA and LORETA methods have been used more than others. Despite the importance of Schizencephaly disease, EEG source localization methods have been less applied in its diagnosis and treatment.

Furthermore, the effects of white matter anisotropic conductivity and the cerebrospinal fluid on the head model of source localization accuracy can be discussed more. Nevertheless, the role of brain source localization algorithms in accelerating diagnosis and treatment of various diseases is quite tangible.

TABLE III
Brain abnormalities and common EEG source localization methods.

| Brain abnormalities | EEG source localization method |
|---|---|
| Meditation | LORETA |
| Conduct problems of adolescents | Current source density-boundary element method |
| Visual attention in the hemi-space | LORETA |
| Subclinical rhythmic electrographic discharges of adults | LORETA |
| Retinotopic organization in the human primary visual cortex | Current source density-boundary element method |
| Type 1 schizencephaly | Electric dipole model |
| Sleep spindles | LORETA and ICA |
| Spider phobia | sLORETA |
| Psychotic symptoms | LORETA |
| To increase working memory | sLORETA |
| OCD | sLORETA |
| Tonic cold pain | sLORETA |
| The primary sensory hand area | Single-dipole localization |
| The phenomena of hypnosis | Minimum-norm method |
| The response inhibition of individuals with psychopathic traits | sLORETA |
| The mechanisms of attention and memory operations | sLORETA |
| Normoacousic tinnitus sufferers | sLORETA |
| The effect of remifentanil drug | sLORETA |
| The Mismatch Negativity (MMN), the P300 and the N400 ERP | LORETA |
| Diabetes mellitus | sLORETA |
| Ischemic brain lesions | Minimum Variance Beamformer |

DISCUSSION AND CONCLUSIONS

Investigating non-invasive source localization approaches can help accelerate the diagnosis and treatment of functional diseases of the brain. Considering the fact that the temporal resolution of the EEG signal is very high compared to other functional methods (such as MRI), it is rather suitable for the evaluation of brain activities during various tasks. Therefore, in this paper systematic review of the research conducted during 1970 to 14 June 2019 regarding the application of the EEG signal in brain source localization as well as its application on the related brain abnormalities is done.

In the selection process of the related studies, databases were searched by using EEG, brain, source localization techniques, and functional abnormalities of the brain keywords. Furthermore, the article selection criteria are considered in this process, such that, at the first stage, their titles and abstracts were investigated; then the articles were selected based on their full-text contents.

As an achievement of this review, it is revealed that the key issues of EEG source localization studies are classified into six important categories, such as solving the inverse problem by statistical methods, improving EEG source localization methods by non-statistical strategies, diagnosis of brain abnormalities using common EEG source localization methods, investigating the effect of head model on EEG source imaging results, detection of epileptic seizures by brain activity localization based on EEG signals and finally, diagnosis and treatment of ADHD abnormalities.

In fact, the main purpose of this study was to answer the aforementioned four research questions in the Introduction section, i.e., RQ1 to RQ4.

Regarding RQ1, it is revealed that more than 42 different statistical method are proposed to localize brain activity sources using EEG signals. Details of each method have been explained in Section 3.1. The disadvantages of the LORETA technique is low spatial resolution and blurred localized images of a point source with a dispersion in the image. The results demonstrate that the principal component analysis is almost useless for isolating spikes and sharp wave activities in an EEG from a patient with epilepsy. On the other hand, it is shown that, in comparison, common spatial pattern performs significantly better.

Moreover, the FOCUSS algorithm has better localization accuracy as compared with other methods and it is able to manage non-uniquely localized energy sources. BPNN is appropriately applied because it has the ability to install an inverse function using training data. But the use of neural networks, in this case, is very limited. As there exist noise and error in the signal subspace and forward model, the selection of the best projection location in the practical case is an important factor. The MUSIC algorithm has several limitations in terms of localizing synchronous sources. A modification of MUSIC algorithm is recursive MUSIC algorithm which can resolve the limitations of MUSIC through the use of spatio-temporal independent topographies model. Localization accuracy of sLORETA and eLORETA methods is higher than LORETA, but their spatial resolution is not appropriate. Also, the cLORETA algorithm works on a surface grid. Followed by sLORETA, the smallest computational complexity belongs to this algorithm. sLORETA and LORETA seem to be the most popular among brain source localization methods. The reasons for this issue can be as follows:

- Both are L2-norm-based solutions and therefore have closed-form expression which makes the computation very efficient. Basically, a Kernel is calculated for the patient only once, and every new source estimate is just the multiplication of the kernel with the measurement.
- Both methods are provided in the form of the toolbox/open source libraries and hence they are easy to be used by a non-technical person [23].

Spatial information about spikes has high correlation with the background signals. This issue leads to the low accuracy of hybridization of ICA and recursively applied and projected multiple signal classification (RAP-MUSIC). Based on the computed results, the advantage of ExSo-MUSIC approach is its higher performance compared to the classical MUSIC algorithms.

In the recent decade, Bayesian methods have become widely used to solve the inverse problem. In 2007, Bayesian methods were used by extracting knowledge about timing and spatial covariance properties of sensor data from evoked sources, interference sources and sensor noise to estimates their contributions. In 2008, a probabilistic approach was proposed that localizes the source activity using a linear mixture of temporal basis functions (TBFs) learned from the data. Evaluation of this method demonstrates significant improvement over existing source localization methods. Considering the two-dipole localization, GA and DE have better performance than SA and PSO methods, but DE needs the setting of some suitable parameter. By reducing the signal-to-noise ratio, the efficiency of all algorithms is decreased, while SA and PSO seemed to be very sensitive to the correlation between sources. The results show that the correlation between sources strongly affects SA and PSO outputs. Generally, among these four methods, GA has better computational cost and performance.

Brain sources have different depths. Deep brain structures play important roles in brain function. For example, brainstem and thalamic relay nuclei have a central role in sensory processing. Many methods fail to deal with deep sources, while minimum norm solution is one of the few methods which works efficiently with sources of different depths.

It is very classical in EEG analysis to consider an additive white Gaussian noise of variance $\delta_n^2$. When this assumption does not hold, it is common to estimate the noise covariance matrix from the data and whiten the data before applying the source localization algorithm. Recently, sparse Bayesian learning algorithm uses an estimate of the sensor noise covariance for brain source localization. In this method, a good initialization of the group-sparsity profile of the sources using brain atlases is applied. Simulations show that the method is robust against measurement noise, while performing faster than the existing methods in real-time circumstances. These results demonstrate that the localization techniques which use brain atlases to localize sources have better performance than the other methods. Each group of the sources is considered in one region of the brain corresponding to the brain atlases. Therefore, the use of brain atlases in the future will greatly increase the localization performance. It is hoped that this point will be taken into consideration in the future work.

In accordance with the statements in the previous sections, the second research question, i.e., RQ2, arises, that is which diseases have been diagnosed and treated by the brain source localization methods so far? From 1995 to 2018, the issue of EEG source localization of interictal spikes has been very much considered. Also, neural mechanisms of ADHD were studied using source localization methods. On the other hand, it is shown that different brain source localization techniques are effective in the diagnosis and treatment of several brain abnormalities and diseases. In this case, performed research is summarized in Section 3.4.3. Among various diseases, ADHD and epilepsy have a greater contribution to the use of EEG source localization techniques. The sources of ictal epileptic activity are well determined by high-resolution EEG and can be used as a basis for epileptic surgery. Comparing ECD, MUSIC, LORETA and sLORETA, ECD approach shows the highest accuracy. The automatic ADHD subtype diagnostic method reduces the need for the skilled psychologists at the diagnostic stage. Therefore, the cost of ADHD diagnosis for families is reduced using automatic and low-cost methods. Furthermore, ADHD disorder diagnosis/treatment at an early age, reduces the government expenses for the treatment of individuals with mental disorders. Finally, by recognizing the difference of the brain regions activity levels in various ADHD patterns, different treatments can be prescribed based on the ADHD pattern of the individual.

Finally, it is known that diabetes mellitus and structural and functional changes occur in the central nervous system. In this case, localization results disclose that the reason for these changes is the frontal region activity of the delta band and the activity of central cortical areas of the gamma band. Also, source activity is decreased in the left postcentral gyrus for the gamma band and in the right superior parietal lobule for the alpha1 (8-10Hz) band.

Turning to the third research question RQ3 about the factors affecting the accuracy of the EEG source imaging methods, it can be said that according to the obtained results, considering the effects of the head cavities, reducing the location error of the electrodes and enough sampling of the potential surface field can improve the localization approaches. Increasing the number of electrodes improves the source localization results, but the absolute enhancement is less considerable for larger electrode numbers. Enough sampling of the potential surface field, a careful conducting volume estimation (head model) and a convenient and well-understood inverse technique are effective factors in the accuracy of EEG source localization.

Finally, as regards the last research question RQ4 about the implications for future research on brain source localization techniques in brain functional abnormalities, it is revealed that from 1995 to 2018, the issue of EEG source localization of interictal spikes has been very much considered. Neural mechanisms of ADHD were studied using source localization methods. sLORETA algorithm has been used more than other localization methods, while new emerging localization methods with better performance and accuracy can be used instead. Different ADHD pattern classification by determining the active regions of the brain in a specific task is an important issue to be considered, but enough research has not been performed so far. Such studies increase the treatment speed and reduce its cost.

The effect of psychiatric drugs on the activity of brain sources has been less considered, hoping to be studied further in the future research using EEG source localization methods. Furthermore, the effects of white matter anisotropic conductivity and the cerebrospinal fluid on the head model of source localization accuracy can be discussed more. The presented findings of this article show that significant developments have been made in the field of brain source localization and will continue. The importance of these methods accuracy is quite clear in diseases such as epilepsy. It is also clear that different psychiatric disorders are more likely to be diagnosed and cured through brain source localization methods. In the future, it is necessary to address the limitations of electroencephalogram-based source localization methods to solve the problems of applied neuroscience.